\documentclass[9pt,twocolumn,twoside]{opticajnl}
\journal{opticajournal}
\setboolean{shortarticle}{true}

\usepackage{lineno}
\usepackage[version=4]{mhchem}

\title{Terahertz electro-optic Kerr effect in \ce{LaAlO_3}}

\author[1*,$\dag$]{Sergey~Kovalev}
\author[1,$\dag$]{Changqing~Zhu}
\author[1]{Anneke~Reinold}
\author[1]{Max~Koch}
\author[1,2]{Zirui~Wang}
\author[1]{Patrick~Pilch}
\author[1]{Ahmed~Ghalgaoui}
\author[1]{Siyu~Duan}
\author[3]{Cong~Li}
\author[3]{Jianbing~Zhang}
\author[3,4]{Pu~Yu}
\author[1]{Zhe~Wang}

\affil[1]{Department of Physics, TU Dortmund University, 44227 Dortmund, Germany}
\affil[2]{School of Physics, Nanjing University, 210093 Nanjing, China}
\affil[3]{State Key Laboratory of Low-Dimensional Quantum Physics, Department of Physics, Tsinghua University, 100084 Beijing, China}
\affil[4]{Frontier Science Center for Quantum Information, 100084 Beijing, China}
\affil[$\dag$]{The authors contributed equally to this work}
\affil[*]{sergey.kovalev@tu-dortmund.de}
\begin{abstract} 

In this letter, we investigate the terahertz (THz) electro-optic Kerr effect (KE) dynamics in \ce{LaAlO_3} (LAO), a widely used substrate for thin film preparation. We show that the KE dynamics strongly depend on the material anisotropy due to interference between THz field-induced and strain-induced optical birefringence. Such interference leads to quasi-phase matching conditions of the KE, which becomes strongly frequency dependent. Depending on the THz frequency, the KE exhibits a uni- and bipolar shape of the quadratic response. The demonstrated effects will be present in a wide variety of materials used as substrates in different THz-pump laser-probe experiments and need to be considered in order
to disentangle the different contributions to the measured ultrafast dynamic signals.

\end{abstract}

\setboolean{displaycopyright}{false}
\doi{}
\dates{}

\begin{document}

\maketitle

The possibility of coherent control of material properties at the light frequency has stimulated extensive research in the field of THz field induced dynamics in condensed matter. The coupling of THz fields with superconducting order parameters \cite{Matsunaga:2014, Hao:2023, Kovalev:2021}, spin polarisation \cite{Ilyakov:2023}, topologically protected surface states \cite{Makushko:2024, Kovalev:2021npj} holds great promise for the next generation of technology.
The transition metal oxides are a broad class of materials where novel functionality at THz frequencies has recently been predicted and partially realised. It has been shown that THz light can control the orientation of the orbital domains in \ce{La_{0.5}Sr_{1.5}MnO_4} \cite{Miller:2015}, THz light-driven multiferroicity has been demonstrated in \ce{SrTiO_3} \cite{Basini:2024}, opening up new prospects for controlling magnetisation by pumping phonons. In addition, the interface between \ce{SrTiO_3} and \ce{LaAlO_3} can host a 2D electron gas that can be controlled by electrical gating \cite{Seo:2024}. The metal oxide interface (\ce{LaAlO_3}/\ce{SrTiO_3}) exhibits various unique physical properties, including the unconventional superconductivity and magnetism \cite{Chen:2024}.  
In order to study novel functionalities of metal oxide based devices on ultrafast time scales, it is necessary to know the different processes taking place in such material heterostructures under THz light illumination. The THz field induced dynamics in \ce{SrTiO_3} has recently been studied, where electronic and ionic Kerr type nonlinearities have been disentangled \cite{Basini:2024prb}. In contrast, the THz driven dynamics in \ce{LaAlO_3}, a common substrate for the oxide electronics, has not yet been explicitly investigated.

\begin{figure}[h!]
\centering
\includegraphics[width=\linewidth]{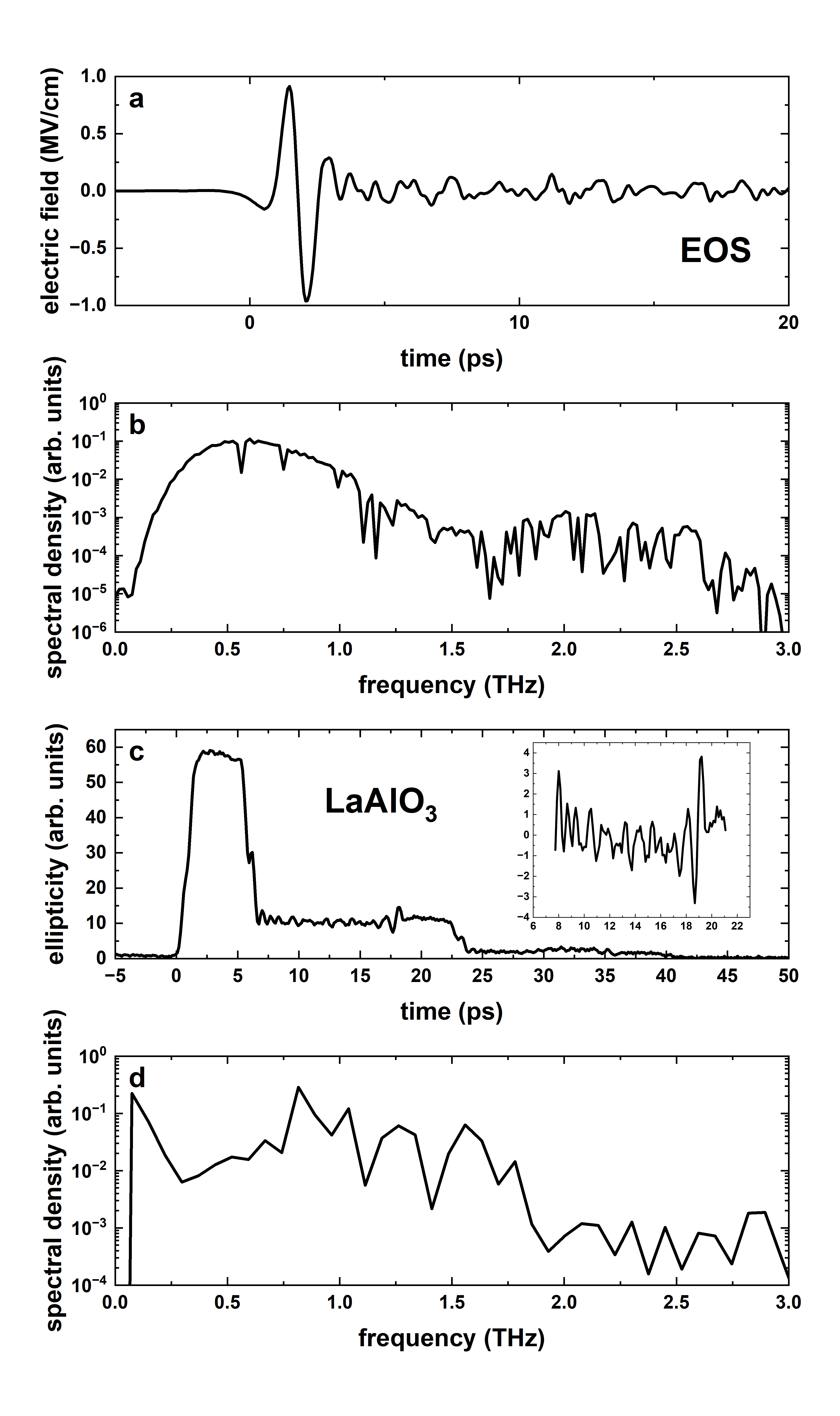}
\caption{\label{Fig:1} Time-domain electro-optical sampling (EOS) of the THz pump pulses measured with 2 mm ZnTe crystal (a) and the corresponding spectrum (b). The THz response measured in the LAO sample as a variation of the probe pulse ellipticity (c). The inset corresponds to the AC filtered signal between 8 and 21 ps, and its spectrum (d).}
\end{figure}

In this Letter we investigate the THz driven dynamics in \ce{LaAlO_3} crystals. We show that the electronic Kerr response is the main contributor to the observed signals and that its shape strongly depends on whether the sample has a single-domain or twin-domain structure. The combined contribution of effects such as phase matching conditions, sample anisotropies at both THz and optical frequencies, and sample birefringence must be taken into account to understand the electronic nonlinear response of the system and thus separate it from other possible THz field induced material phases.

The experiment was performed with 3~mJ laser pulses and is similar to that described in Ref.~\cite{Kovalev:2024}. Using a tilted pulse front generation scheme in a \ce{LiNbO_3} prism, THz pulses with up to 900~kV/cm peak field strength irradiated the samples, whose response was probed by a subsequent femtosecond laser pulse.
Additional THz bandpass filters were used to generate narrowband THz pulses \cite{Kovalev:2021}. 
The THz pulse induced ellipticity and the polarisation rotation of the probe pulse were measured as a function of the delay between the THz pump and the optical probe pulses. 
As cubic perovskites with lattice parameter a = 3.792 \r{A}, 0.5 mm thick \ce{LaAlO_3} single crystals were synthesised by the Czochralski process and cut perpendicular to the [001] axis.
Mechanical treatment (sample cutting and polishing) of \ce{LaAlO_3} samples characterised by optical anisotropy as a result of twin domain formation  \cite{Salje:2016}. The mechanically treated samples are further referred  to as strained samples. 
The unstrained (single-domain) samples are free optical anisotropy.

Figure \ref{Fig:1}a shows a temporal profile of the THz field at the sample position, measured by electro-optical sampling using a 2~mm ZnTe crystal. The corresponding spectrum of the THz pulse is shown in Fig.~\ref{Fig:1}b. The peak THz field reached the value of 900~kV/cm and its spectrum was concentrated in the frequency range between 0.3 and 1 THz. The THz driven dynamic in LAO is probed by the ellipticity change of the 800 nm probe pulse, and is shown in Fig.~\ref{Fig:1}c.
The measured signals have a unipolar shape with  their amplitude scales quadratically with the THz field strength.
The signal in the time window from 0 ps to 6.5 ps with a FWHM of about 5 ps can be attributed to the overlap between copropagating THz and laser pulses in the LAO sample. 
With collinear propagation of the crystal thickness $L$, the THz and optical pulses are separated in time by $\delta \tau_1 = \frac{L}{c} \left( n_{THz}- n_{opt}\right)$, resulting in a broadening of the measured signal. The refractive index of LAO at THz frequencies $n_{THz}$ is about 5 \cite{Hughes:2014}, while at 800 nm $n_{opt}=2$ \cite{Rizwan:2019}. These values correspond to $\delta \tau_1$ = 5 ps for 0.5 mm thick LAO, which is in good agreement with the experimental data.
The measured signal between 6.5 ps and 18.5 ps arises from the interaction between the probe pulse and the THz pulse reflected from the back surface of the sample (the optical and THz pulses propagate in opposite directions). In this configuration the signal width $\delta \tau_2 = \frac{L}{c} n_{THz}$ is about 8.5 ps for 0.5 mm of LAO. The large velocity mismatch between the THz and optical pulses does not allow to detect the high frequency content of the THz quadratic signals.
At the same time, in the case of THz reflection at the boundary, the interference between the incident and reflected pulses will mitigate such requirements, and high frequency THz content can be observed, seen as oscillations between 6.5 ps and 18.5 ps. The inset of Fig.~\ref{Fig:1}c corresponds to the THz oscillatory part. The shape of the oscillatory part consists of two  counterpropagating wavepackages, corresponding to the first reflection from the back surface (starting at 6.5 ps) and to the second reflection from the front surface (starting at 19 ps and propagating in the reverse direction).
The spectrum of the THz oscillatory part (Fig.~\ref{Fig:1}d) is broad and has a frequency content at the second harmonic frequency of the incident THz pump pulse, ranging from 0.75 up to 1.7 THz. At 18.5 ps time delay, the back propagating THz pulse is reflected from the front surface and afterwards propagates collinear with the optical probe pulse (18.5 - 23.5 ps time window). In this collinear regime we do not observe any THz oscillations and its width is similar to that of the main signal. 
An interesting observation is that the THz induced signals from the first (6.5 - 18.5 ps) and second reflections (18.5 - 23.5 ps) are very similar in magnitude. This is due to the fact that in the LAO sample the THz reflection losses are compensated by the group velocity mismatching (GVM) for collinear and anticollinear propagation: $\frac{\left(n_{THz}-n_{opt} \right)}{\left(n_{THz}+n_{opt} \right)}$ is the ratio between the GVM from collinear and anticollinear propagation of THz and optical pulses, which is equal 0.43. The THz reflection losses at the LAO boundary are $\left( \frac{n_{THz}-1}{n_{THz}+1} \right)^2 = 0.44$. 
Measurements of the THz induced probe pulse polarisation rotation have not provided any signals above the noise floor, indicating that the THz pulse modulates only the refractive index without inducing sample magnetisation or absorption anisotropy. This suggests that the KE is the source of the observed THz field driven dynamics, as probed by the probe pulse ellipticity modulation.

\begin{figure}[ht]
\centering
\includegraphics[width=\linewidth]{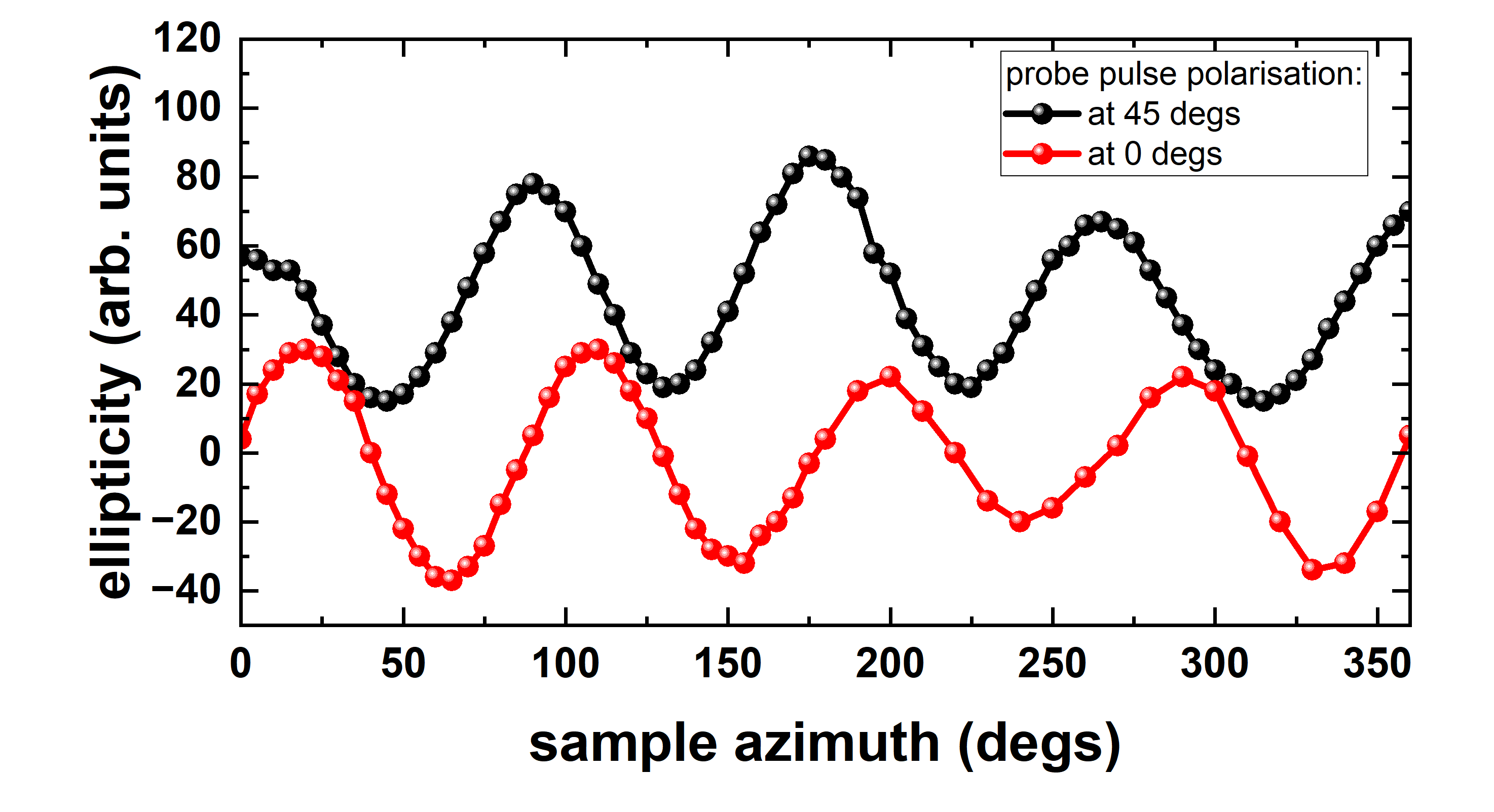}
\caption{\label{Fig:2} The THz field induced probe pulse ellipticity change in unstrained LAO as a function of sample azimuth angle. The THz pulse polarisation is vertical, the probe pulse polarisation is horizontal (0$^\circ$, red curve) and at 45$^\circ$ (black curve).}
\end{figure}

Figure \ref{Fig:2} shows the dependence of the THz Kerr effect in LAO on the sample azimuth angle, performed in two configurations: THz pulse polarisation is orthogonal to the probe pulse polarisation (red curve) and at 45$^\circ$ with respect to each other (black curve). In contrast to the amorphous systems, where the Kerr signal can be measured when the probe pulse polarisation does not coincide with the THz polarisation \cite{Kovalev:2024}, the Kerr signals here have more complex dynamics. For the horizontal probe pulse polarisation, the Kerr signal is proportional to the $\sin \left( 4 \theta\right)$, where $\theta$ is the angle between the $\left[ 100 \right]$ sample axis and the vertical direction. When the probe pulse is at 45$^\circ$, the Kerr signal is proportional to the $1+0.5\cos \left( 4 \theta\right)$. This dependence is due to the presence of several components of the THz nonlinear susceptibility tensor $\chi^{\left( 3 \right)}$, including off-diagonal elements. The symmetry of the THz Kerr response in \ce{SrTiO_3} has recently been analysed, and a similar fourfold symmetry has been attributed to the $\chi^{\left( 3 \right)}_{iijj}$ component \cite{Basini:2024prb}.

\begin{figure}[ht]
\centering
\includegraphics[width=\linewidth]{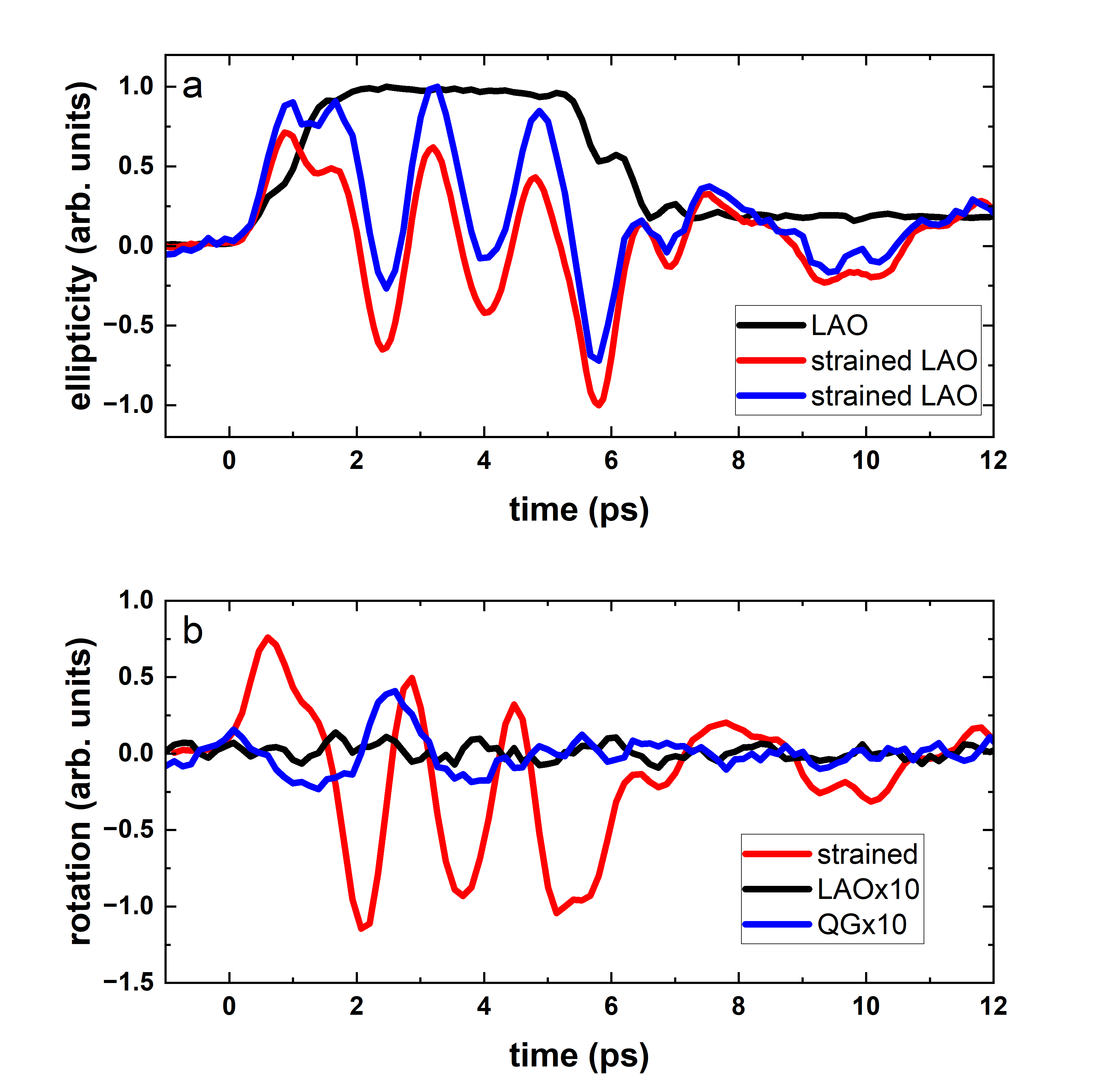}
\caption{\label{Fig:3} THz induced probe beam ellipticity in unstrained and strained LAO samples (a). The red and blue curves correspond to the signals measured in two areas of the same sample. (b) THz induced probe beam polarisation rotation measured in strained LAO, LAO and quartz glass (QG).}
\end{figure}

In Figure~\ref{Fig:3}a we show the probe pulse ellipticity induced in unstrained and strained LAO samples. Both samples had the same growth conditions, but nevertheless their Kerr response is qualitatively very different: besides the unipolar Kerr signal with a hat shape (for the LAO sample, black curve), the strained sample shows an oscillatory behaviour.
The observed unipolar and oscillatory signals scale quadratically with the THz field strength. Moreover, the relationship between these two contributions is inhomogeneous over the sample, as we see the signal changes its shape depending on the sample position (red and blue curves). The duration of these contributions is the same for both samples, about 6 ps, indicating that they originate from the instantaneous Kerr type response and their duration is governed by the GVM between THz and optical pulses.
In addition to the probe pulse ellipticity, the THz pulse also induces the probe pulse polarisation rotation in the strained sample. In Fig.~\ref{Fig:3}b we compare the probe pulse polarisation rotation in LAO, quartz glass (QG) and strained LAO samples. We have observed probe pulse polarisation rotation in QG due to the magneto-optical sampling in Faraday geometry \cite{Kovalev:2024}. At the same time, the polarisation rotation signals from unstrained LAO are not visible. The polarisation rotation in the strained sample is several ten times larger than in QG, and has an oscillatory behaviour similar to the ellipticity dynamics.
{
The dependence of the THz induced rotation of the optical beam polarisation on the azimuth angle of the strained sample is shown in the Supplementary Information.}
The presence of an ultrafast electronic Kerr response, which mitigats the GVM frequency cut-off, was previously demonstrated in \ce{SrTiO_3}. In that case, the strong THz absorption on soft phonons resulted in very short THz penetration in the sample (few $\mu$m), which led to the phase mismatch compensation between propagating THz and probing optical pulses \cite{Basini:2024prb}. Such a scenario is not valid for LAO, while there are no phonon modes at the frequencies used in this work (below 1.5 THz), and the THz absorption, measured with THz time-domain spectroscopy, is small (below 10 cm$^{-1}$, see {
the Supplementary Information}). Furthermore, the THz Kerr signal oscillating in time can have both positive and negative values (bipolar function, strained LAO), whereas the intrinsic Kerr type nonlinearity is a unipolar function (LAO sample) \cite{Basini:2024prb}. 


\begin{figure}[h!]
\centering
\includegraphics[width=\linewidth]{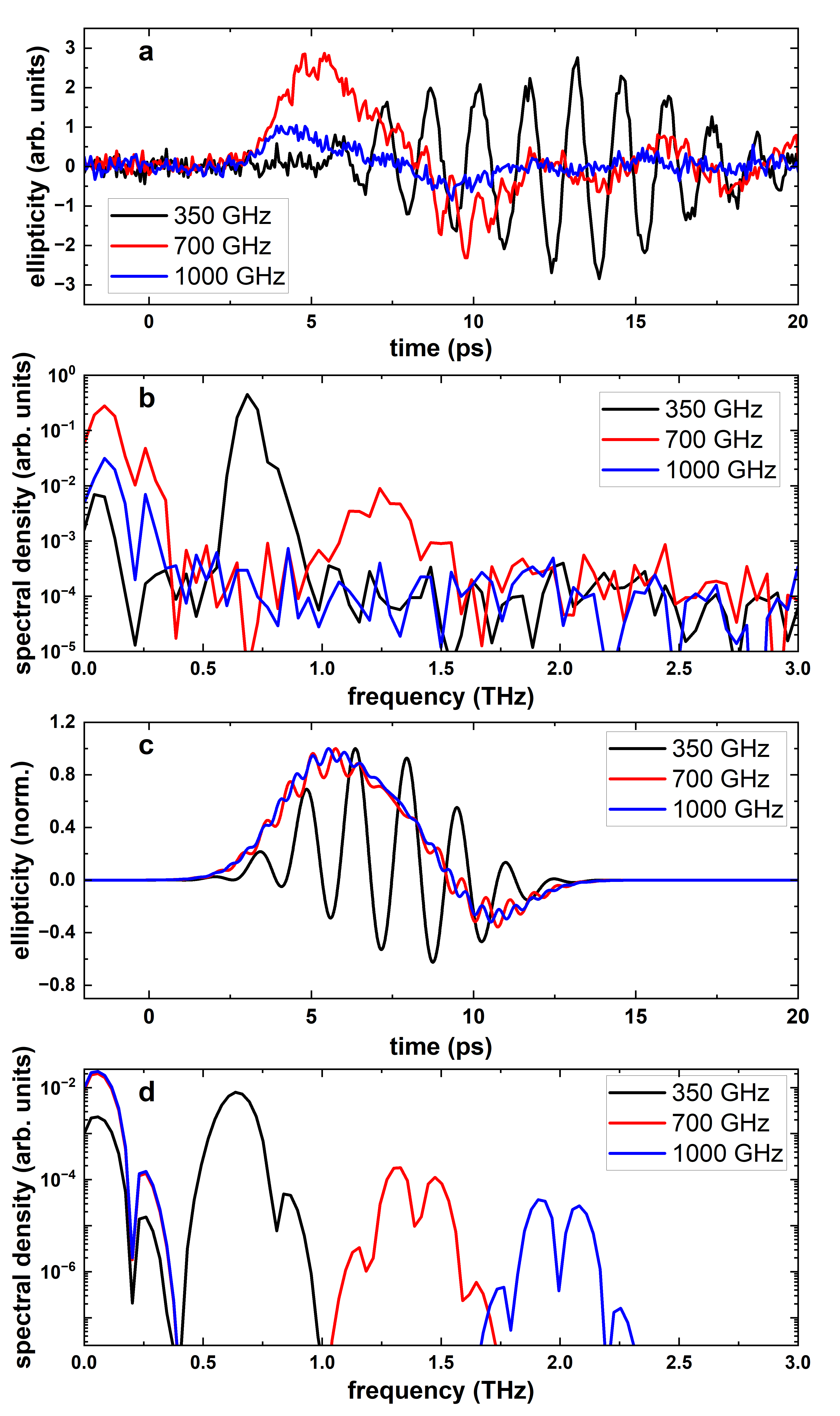}
\caption{\label{Fig:5} Time-domain dynamics of the Kerr signal induced by THz pulses at 350, 700 and 1000 GHz central frequency in a strained LAO sample (a) and the corresponding spectrum (b). The simulations of Kerr induced dynamics in material with optical birefringence under different frequencies of THz pump pulses in time (c) and frequency domain (d).}
\end{figure}

To understand the origin of the unusual Kerr response, we measured the THz induced dynamics in the strained sample using narrowband THz pulses with 20 \% bandwidth and 350 GHz, 700 GHz and 1000 GHz centre frequencies. The corresponding time domain measurements are shown in Fig.~\ref{Fig:5}a and their spectra are shown in Fig.~\ref{Fig:5}b. The signals can be decomposed into two contributions: a unipolar function with a width of about 5 ps (quasi-DC), and an oscillatory part at the second harmonic of the incident pulse frequency. For the pump pulses of 350 GHz, the oscillatory part (700 GHz) is much larger than the quasi-DC contribution, and the total signal looks like a sinusoidal function. At 700 and 1000 GHz the quasi-DC component is much larger than the second harmonic oscillations (at 1000 GHz frequency the oscillatory dynamics are below our detection noise), and the quasi-DC component slowly (within 5 ps) changes sign. 
{
In the strained samples we cannot resolve any THz dynamics at the THz pump pulse frequencies (Fig.~\ref{Fig:5}b). In the single crystal LAO we did not observe any THz signals scaling linearly with the THz field strength, suggesting that the THz Pockels effect in LAO is negligible and the dynamics are governed by second-order processes. This may be due to the cubic structure of LAO, which differs from, for example, crystal quartz (trigonal structure), where both linear and quadratic responses can exist simultaneously \cite{Li:2024}.}

We attribute the observed Kerr response in strained LAO to the interference between strain induced optical anisotropy and the electronic KE. LAO crystals have a pseudocubic lattice structure and are optically isotropic. The investigated unstrained LAO sample did not exhibit any change in the transient polarisation state of the probe pulse. In contrast, the strained sample showed a strong modification of the polarisation state of the probe pulse during passage, indicating a high optical birefringence {
(see the Supplementary Information)}.
Optical birefringence in mechanically stressed LAO has been observed previously and attributed to the appearance of polar tweed domains \cite{Salje:2016}. The optical birefringence causes the probe pulse polarisation rotation and ellipticity to change in the sample. Due to the large GVM in LAO, the THz and probe pulse polarisation will be time dependent at the overlap.
To verify this, we have performed simulations of the measured signal (Fig.~\ref{Fig:5}c and \ref{Fig:5}d). The Kerr response was calculated taking into account the propagation effect in LAO together with the optical birefringence of the probe pulse. In this case, the Kerr response can be approximated as follows {
(see the Supplementary Information)}:
\begin{equation*}
S\left( \tau \right) \sim \int_T E^2_{THz} \left( t+\tau \right) \left[ A + \sin \left( 2 \pi \kappa t \right)\right] dt .
\end{equation*}
Here, the first term in the integral corresponds to the electronic Kerr response, which is instantaneous in time and quadratic in the THz field $E_{THz}$. The second term in the square brackets corresponds to the symmetry of the Kerr response (Fig.~\ref{Fig:2}) and is similar to that calculated for \ce{SrTiO_3} with the fit parameters $A$ and $\kappa$. The integration window $T=5$ ps corresponds to the separation time between optical and THz pulses passing through a LAO sample. The simulated results with $\kappa = 0.7$ ps$^{-1}$ gave similar results to the experiment. At this value, the probe pulse phase retardation is about $4\times 2\pi$ over the propagating 0.5 mm LAO sample, which corresponds to the sample birefringence of about $6 \times 10^{-3}$. This value is of a similar order to that of crystalline quartz and agrees well with previous studies of optical birefringence in strained LAO \cite{Salje:2016}.   
The optical birefringence of the sample could be the reason why we observed additional probe pulse polarisation rotation in the Kerr signal, whereas elliptically polarised pulses will undergo polarisation rotation under KE.
In our simulations we used a simplified model that does not take into account the birefringence at the THz frequencies, the phase response of the THz filters (responsible for the time delay of the different THz components as seen in Fig.~\ref{Fig:5}a) and the possible modification of the crystallographic axis in the domain structures. For a more accurate understanding of the system evolution, these effects must also be considered.
{
The simulations correlate well with the experimental results on the observations of the DC and second harmonic contributions, and relative suppression of the second harmonic signal with excitation frequency.
The DC and second harmonic signals can be interpreted as intrapulse different and sum frequency generation and arise due to the second-order nature of the electro-optic Kerr effect. The ratio between these two processes is determined by the phase matching conditions and hence the $\kappa$ value.
}

In conclusion, we have demonstrated THz field induced Kerr dynamics in single and twin-domain LAO samples. We showed that the sample anisotropy has a strong influence on the Kerr effect dynamics. The combination of effects such as optical anisotropy, group velocity mismatch and the symmetry of the Kerr nonlinearity tensor allows to mitigate the phase mismatch conditions and to observe/simulate the high frequency electronic Kerr response in bulk samples. The observed effects needs be accounted to disentangle different contributions from the ultrafast dynamics in novel systems and their heterostructures.

\begin{backmatter}
\bmsection{Acknowledgments} We acknowledge support by the European Research Council (ERC) under the Horizon 2020 research and innovation programme, grant agreement No. 
950560 and partial support by MERCUR (Mercator Research Center Ruhr) via Project No. Ko-2021-0027.

\bmsection{Disclosures} The authors declare no conflicts of interest.

\bmsection{Data Availability Statement}  Data underlying the results presented in this paper are
 not publicly available at this time but may be obtained from the authors upon
 reasonable request. 
 
\bmsection{Supplemental document} See Supplement 1 for supporting content
\end{backmatter}
\bibliography{MainText}

\end{document}